\documentstyle[aps,prb,epsf]{revtex} 

\newcommand{\subscrpt}[2]{{#1}_{{#2}}}

\newcommand{\bfc}{{\bf c}}
\newcommand{\bfcsub}[1]{\subscrpt{\bfc}{#1}}
\newcommand{\bfci}{\bfcsub{i}}

\newcommand{\bfx}{{\bf x}}

\newcommand{\bfxt}{(\bfx, t)}
\newcommand{\bfy}{{\bf y}}

\def\bfsigma{\mbox{\boldmath $\sigma$}}
\def\bfomega{\mbox{\boldmath $\Omega$}}

\newcommand{\bge}{\begin{equation}}
\newcommand{\ee}{\end{equation}}
\newcommand{\bgc}{\begin{center}}
\newcommand{\ec}{\end{center}}
\newcommand{\bgea}{\begin{eqnarray}}
\newcommand{\eea}{\end{eqnarray}}
\newcommand{\bgeas}{\begin{eqnarray*}}
\newcommand{\eeas}{\end{eqnarray*}}

\newcommand{\dt}{\Delta t}

\newcommand{\hcc}{H_{\mbox{\scriptsize cc}}}
\newcommand{\hcd}{H_{\mbox{\scriptsize cd}}}
\newcommand{\hdc}{H_{\mbox{\scriptsize dc}}}
\newcommand{\hdd}{H_{\mbox{\scriptsize dd}}}

\begin{document}

\title{
\begin{flushleft}
{\footnotesize OUTP-9612S}\\
{\footnotesize BU-CCS-960201}\\[0.3in]
\end{flushleft}
{\bf Lattice-Gas Simulations of Domain Growth, Saturation and Self-Assembly
     in Immiscible Fluids and Microemulsions}
}
\author{
Andrew N. Emerton\\
{\small \sl Department of Theoretical Physics,}\\
{\small \sl Oxford University,}\\
{\small \sl 1 Keble Road, Oxford OX1 3NP, U.K.}\\
{\small \tt emerton@thphys.ox.ac.uk}\\
Peter V. Coveney\\
{\small \sl Schlumberger Cambridge Research,}\\
{\small \sl High Cross, Madingley Road, Cambridge CB3 0EL, UK.}\\
{\small \tt coveney@cambridge.scr.slb.com}\\
Bruce M. Boghosian\\
{\small \sl Center for Computational Science, Boston University,}\\
{\small \sl 3 Cummington Street, Boston, Massachusetts 02215, U.S.A.} \\
{\small \tt bruceb@bu.edu}\\
[0.3cm]
}
\date{\today}
\maketitle

\begin{abstract}
  We investigate the dynamical behavior of both binary fluid and
  ternary microemulsion systems in two dimensions using a recently 
  introduced hydrodynamic
  lattice-gas model of microemulsions. We find that the presence of
  amphiphile in our simulations reduces the usual oil-water interfacial
  tension in accord with experiment and consequently affects the
  non-equilibrium growth of oil and water domains. As the density of
  surfactant is increased we observe a crossover from the usual
  two-dimensional binary fluid scaling laws to a growth that is {\it
  slow}, and we find that this slow growth can be characterized by a
  logarithmic time scale. With sufficient surfactant in the system we
  observe that the domains cease to grow beyond a certain point and we
  find that this final characteristic domain size is inversely
  proportional to the interfacial surfactant concentration in the
  system.
\end{abstract}

\section{Introduction}
The introduction of amphiphilic molecules into a system of oil and water
is known to have marked effects on the properties and behavior of such
mixtures. As a result of the particular physical and chemical properties
of surfactant molecules one can observe the formation of a wealth of
complex structures. For a general review see Gelbart {\it et
al.}~\cite{bib:gea}. One major feature of these systems is that the
usual oil-water interfacial tension is dramatically lowered by the
presence of amphiphile~\cite{bib:gas}, this being the origin of much of
the commercial interest in such self-assembling structures.  In this
paper we demonstrate that our recently introduced hydrodynamic
lattice-gas model of microemulsions~\cite{bib:bce} is able to replicate
this important experimentally observed phenomenon.

Growth kinetics in binary immiscible fluids have received much attention
recently. Phase separation in these systems has been simulated using a
variety of techniques: these include cell dynamical systems without
hydrodynamics~\cite{bib:so} and with Oseen tensor
hydrodynamics~\cite{bib:so2}; time-dependent Ginzburg-Landau models
without hydrodynamics~\cite{bib:ctg}, and
with hydrodynamics~\cite{bib:fv,bib:vf,bib:walc}; as well as lattice-gas
automata~\cite{bib:bal,bib:rothmanrefs} and the related
lattice-Boltzmann techniques~\cite{bib:acg}. A central quantity in the
study of growth kinetics is the time-dependent average domain size
$R(t)$. For binary systems in the regime of sharp domain walls, this
follows algebraic growth laws of the form $R(t) \sim t^{n}$. In general,
previous simulations have confirmed experimental observations and
theoretical predictions~\cite{bib:bray,bib:sed} for these systems. That
is, for models without hydrodynamic interactions (binary alloys) the
growth exponent is found to be $n = \frac{1}{3}$, independent of the
spatial dimension. If flow effects are relevant (binary fluids), and the
domain size $R$ is greater than the hydrodynamic length $R_{h} =
\frac{{\nu}^2}{\rho \sigma}$~\cite{bib:bray}, where $\nu$ is the
kinematic viscoscity, $\rho$ is the density and $\sigma$ is the surface
tension coefficient, then one obtains $n = \frac{2}{3}$ in two space
dimensions. We use our lattice-gas model to investigate this as well as
the less commonly observed $R < R_{h}$ regime in $2D$; this is described
in Sec.~\ref{sec:bif}. In three dimensions in the regime $R < R_{h}$
the growth exponent is $n = \frac{1}{3}$ crossing over to $n = 1$ at
late times, with $n = \frac{2}{3}$ if $R > R_{h}$.

The lowering of the oil-water interfacial tension by amphiphile has
important consequences for the non-equilibrium dynamical growth of
domains within ternary systems.  Our microscopic lattice-gas model of
microemulsions, which correctly models the mesoscopic and macroscopic
fluid behavior, enables us to investigate such dynamical domain growth.
In ternary systems, growth kinetics has previously been studied by
numerical integration of time-dependent Landau-Ginzburg models, for
example the hybrid model of Kawakatsu {\it et al.}~\cite{bib:klsd} and
the two local order parameter model of Laradji {\it et
  al.}~\cite{bib:lggz,bib:lhgz}. These models do not include
hydrodynamic effects and find that surfactants modify the dynamics from
the binary $n = \frac{1}{3}$ algebraic exponent to a slow growth that
may be logarithmic in time. More recently Laradji {\it et
  al.}~\cite{bib:lmtz} have modelled phase separation in the presence of
surfactants using a very simple molecular dynamics model which
implicitly includes hydrodynamic forces. These authors found that such
systems exhibit nonalgebraic, slow growth dynamics and that the average
domain size saturates at a value inversely proportional to the
surfactant concentration. They also found a crossover scaling form which
describes the change from the algebraic growth in pure binary fluids to
a slower domain growth when surfactants are present. This scaling {\em
  ansatz} was observed to hold with an exponent $n = \frac{1}{2}$, which
corresponds to the binary growth exponent seen at intermediate times in
$2D$ in other recent simulations, as discussed in Sec.~\ref{sec:bif}.
Contrasting with these results, Patzold and Dawson~\cite{bib:pad} have
recently suggested that with noise included in a time-dependent
hydrodynamic Ginzburg-Landau model, binary-fluid-like power law growth
behavior can be observed across the whole range of surfactant
densities, with the exponent decreasing as the amount of surfactant is
increased. However, this is clearly not how real microemulsion systems
behave: Even with noise present in the system, one would expect domain
growth to cease once sufficient surfactant is present, and consequently
one would also expect the growth to slow down significantly prior to
this. We believe that the model employed by Patzold and Dawson has not
allowed them to access the true late-time dynamics and that the scaling
region within which they calculated exponents is probably too short for
intermediate to high surfactant concentrations. Access to this
asymptotic regime is also difficult for molecular dynamics simulations.
Lattice-gas automaton models, on the other hand, while including by
construction the correct hydrodynamics, permit simulations over a wider
range of relevant time scales than those systems described above.
Fluctuations are an inherent and important physical component of such
models, and consequently our microemulsion model provides useful and
arguably unique insight into the dynamics of such complex systems.

Characterization of the slow growth found in these amphiphilic systems
prior to saturation of the domain size remains a challenge. Comparative
slowing down in domain growth has been observed in $2D$ simulations of
systems with quenched impurities~\cite{bib:ghss}, where the growth is
described by a logarithmically slow activated process with $R(t) \sim
(\ln t)^{\theta}$. For these systems the surface tension diminishes over
time, whereas for amphiphilic systems, which may not be quenched in the
same sense as the impurities in these models, the surface tension begins
to be affected as soon as the molecules reach oil-water interfaces. Here
we are interested in the growth laws that are observed as we approach
the ``saturation'' point in our system. This is the point at which
self-similarity and the usual scaling laws must break down: In some
sense there must be an exponential tail-off in the growth of domain size
as the saturation point is reached.  The results, obtained using our
microemulsion model, for domain growth in these ternary systems as the
quantity of surfactant is varied are presented in Sec.~\ref{sec:sadm}.

\section{The Lattice-Gas Automaton Model}

Our lattice-gas model is based on a microscopic particulate format that
allows us to include dipolar surfactant molecules alongside the basic
oil and water particles~\cite{bib:bce}. In this paper we are concerned
only with a two-dimensional version of the model, though an extension to
$3D$ is currently underway~\cite{bib:toappear}. Working on a triangular
lattice with lattice vectors $\bfci$ ($i=1,\ldots,6$), the state of the
$2D$ model at site $\bfx$ and time $t$ is completely specified by the
occupation numbers $n_i^{\alpha}(\bfx,t)\in\{0,1\}$ for particles of
species $\alpha$ and velocity $(\bfci/ \Delta t)$.

The evolution of the lattice gas for one timestep takes place in two
substeps.  In the {\it propagation} substep the particles simply move
along their corresponding lattice vectors.  In the {\it collision}
substep the newly arrived particles change their state in a manner that
conserves the mass of each species as well as the total $D$-dimensional
momentum.

We allow for two immiscible species which, following convention, we
often represent by colors: $\alpha=B$ (blue) for water, and $\alpha=R$
(red) for oil, and we define the {\it color charge} of a particle moving
in direction $i$ at position $\bfx$ at time $t$ as $q_i\bfxt\equiv
n_i^R\bfxt-n_i^B\bfxt$.  Interaction energies between outgoing particles
and the total color charge at neighbouring sites can then be calculated
by assuming that a color charge induces a {\it color potential} $\phi
(r)= q f(r)$, at a distance $r$ away from it, where $f(r)$ is some
function defining the type and strength of the potential.

To extend this model to amphiphilic systems, we also introduce a third
(surfactant) species $S$, and the associated occupation number
$n_i^S\bfxt$, to represent the presence or absence of a surfactant
particle.  Pursuing the electrostatic analogy, the surfactant particles,
which generally consist of a hydrophilic portion attached to a
hydrophobic (hydrocarbon) portion, are modelled as {\it color dipole
  vectors}, $\bfsigma_i\bfxt$. As a result, the three-component model
includes three additional interaction terms, namely the color-dipolar
field, the dipole-color field and the dipole-dipole interactions.

The total interaction energy that results can be written
\bge
\Delta H_{\mbox{\scriptsize int}}
    =  \Delta \hcc +
         \Delta \hcd +
         \Delta \hdc +
         \Delta \hdd \\
    =  \left[
           \left(
             {\bf J} +
             \frac{\bfsigma^\prime}{\dt}
           \right)\cdot
           \left(
             {\bf E} + {\bf P}
           \right) +
           {\cal J} :
           \left({\cal E} + {\cal P}\right)
         \right]\dt.
\label{eq:tie}
\ee
where we have defined the {\it color flux} of an outgoing state
\bge
{\bf J}\bfxt\equiv \sum_i^n \frac{\bfci}{\dt} q_i^\prime\bfxt
\label{eq:cf}
\ee
(the sum extending over all lattice vectors at a site, so that in
this case $n = 6$), and the {\it color field}
\bge
{\bf E}\bfxt\equiv \sum_{\bfy\in {\cal L}} f_1(y)\bfy q(\bfx + \bfy,t),
\label{eq:cff}
\ee
where the sum is over sites ${\bf y}$ which are elements of the
hexagonal lattice ${\cal L}$.  For short-range forces, the function
$f_1$ has compact support so that this sum includes only sites nearby
$\bfx$.  The {\it dipolar field} vector is
\bge
{\bf P}\bfxt\equiv
   -\sum_{\bfy\in {\cal L}}
   \left[
   f_2(y)\bfy\bfy -
   f_1(y) {\bf 1}
   \right]\cdot\bfsigma (\bfx+\bfy,t),
   \label{eq:dfv}
\ee
where ${\bf 1}$ denotes the rank-two unit tensor and
$\bfsigma^\prime\equiv\sum_i\bfsigma_i$ represents the total outgoing
dipolar vector at a site.  Similarly we have defined the {\it dipolar
  flux tensor}
\bge
{\cal J}\bfxt\equiv
   \sum_i^n\frac{\bfci}{\Delta t}\bfsigma_i^\prime\bfxt,
   \label{eq:df}
\ee
and the {\it color field gradient tensor}
\bge
{\cal E}\bfxt\equiv
   \sum_{\bfy\in {\cal L}} q(\bfx+\bfy,t)
   \left[
   f_2(y)\bfy\bfy -
   f_1(y) {\bf 1}
   \right],
   \label{eq:cfg}
\ee
where again ${\bf 1}$ denotes the rank-two unit tensor. Finally we
have the {\it dipolar field gradient} tensor
\bge
{\cal P}\bfxt =
   -\sum_{\bfy\in {\cal L}}
   \bfsigma (\bfx+\bfy, t)\cdot
   \left[
   f_3(y)\bfy\bfy\bfy -
   f_2(y)\bfy\cdot\bfomega
   \right],
   \label{eq:dfg}
\ee
wherein $\bfomega$ is the completely symmetric and isotropic fourth-rank
tensor. In Eqs.~(\ref{eq:cff}), (\ref{eq:dfv}), (\ref{eq:cfg}) and
(\ref{eq:dfg}) we have defined certain derivatives of the function
$f(r)$
\bge
f_\ell (y)\equiv\left(-\frac{1}{y}\frac{d}{dy}\right)^\ell f(y),
\label{eq:fdef}
\ee
where $\ell$ is a positive integer or zero~\cite{bib:bce}.

The collision process of the algorithm consists of enumerating the
outgoing states allowed by the conservation laws, calculating the total
interaction energy for each of these, and then, following the ideas of
Chan and Liang~\cite{bib:cal} (see also Chen {\em et al.}~\cite{bib:chen}),
forming Boltzmann weights
\bge
e^{-\beta \Delta H},
\label{eq:bd}
\ee
where $\beta$ is an inverse temperature-like parameter.  The
post-collisional outgoing state and dipolar orientations can then be
obtained by sampling from the probability distribution formed from these
Boltzmann weights; consequently the update is a stochastic process. The
dipolar orientation streams with surfactant particles in the usual way.

Our model's parameter space has certain important pairwise limits. With
no surfactant in the system, Eq.~(\ref{eq:tie}) reduces to the
color-color interaction term only,
\[
\Delta \hcc \equiv  {\bf J} \cdot {\bf E},
\]
which we note to be exactly identical to the expression for the total
color work used by Rothman and Keller~\cite{bib:rk} to model immiscible
fluids.  Correspondingly, with no oil in the system we are free to
investigate the formation and dynamics of the structures that are known
to form in binary water-surfactant solutions.  Indeed, in our original
paper~\cite{bib:bce} we investigated both of these limits.  In the limit
of no surfactant we obtained immiscible fluid behavior similar to that
observed by Rothman and Keller, and for the case of no oil in the system
we found evidence for the existence of micelles and for a critical
micelle concentration.  Moreover, we demonstrated that this model
exhibits the correct $2D$ equilibrium microemulsion phenomenology for
both binary and ternary phase systems using a combination of visual and
analytic techniques; various experimentally observed self-assembling
structures, such as the droplet and bicontinuous microemulsion phases,
form in a consistent manner as a result of adjusting the relative
amounts of oil, water and amphiphile in the system. The presence of
enough surfactant in the system is shown to halt the expected phase
separation of oil and water, and this is achieved without altering the
coupling constants from values that produce immiscible behavior in the
case of no surfactant.

Note that in order to incorporate the most general form of interaction
energy within our model system, we introduce a set of coupling constants
$\alpha, \mu, \epsilon, \zeta$, in terms of which the total interaction
energy can be written as
\bge
\Delta H_{\mbox{\scriptsize int}}
       =   \alpha \Delta \hcc +
         \mu \Delta \hcd +
         \epsilon \Delta \hdc +
         \zeta \Delta \hdd.
\label{eq:tiw}
\ee
These terms correspond, respectively, to the relative
immiscibility of oil
and water, the tendency of surrounding dipoles to bend round oil or
water particles and clusters, the propensity of surfactant molecules to
align across oil-water interfaces and a contribution from pairwise
(alignment) interactions between surfactants.  In the present paper we
analyze domain growth of critical quenches within both binary and
ternary systems and consequently the two coefficients with which we are
most concerned are $\alpha$ and $\epsilon$.

\section{Surface Tension Analysis}
\label{sec:sta}

The lowering of the interfacial tension between oil and water by the
action of surfactant molecules located at such interfaces is an
important property of microemulsions. Experimental investigation of
polymer/block-copolymer systems, where the timescales are much slower
than in the related microemulsions, has made studies of such
characteristics possible~\cite{bib:agk}.  In the present section, we
analyze the surface tension within a system of oil, water and surfactant
as it varies with the surfactant density (concentration).  We work with
$\beta = 1.0$ and use
\bgeas
  \alpha &=& 1.0  \\
     \mu &=& 0.001\\
\epsilon &=& 8.0  \\
   \zeta &=& 0.005
\eeas
as the values of the coefficients in Eq.~(\ref{eq:tiw}), strongly
encouraging surfactant molecules to accumulate at oil-water interfaces
while maintaining the normal oil-water immiscible behavior.

\begin{figure}
\begin{center}
\leavevmode
\hbox{%
\epsfxsize=3.5in
\epsffile{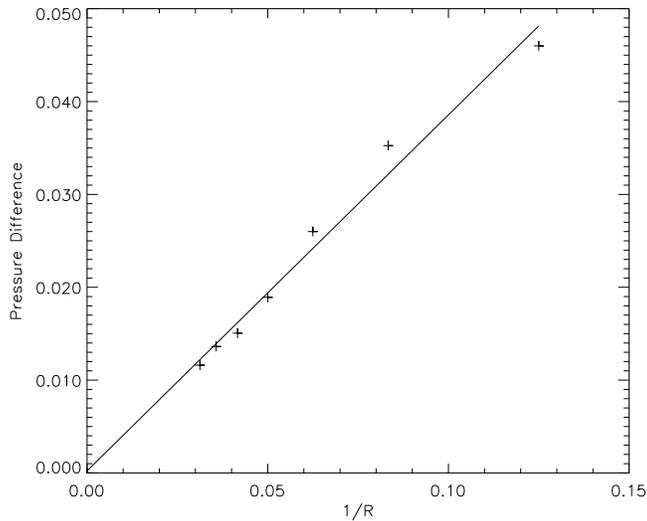}}
\end{center}
\caption{Verification of Laplace's law and estimation of surface
tension for two immiscible fluids (oil and water) only.}
\label{fig:stb}
\end{figure}

We begin by showing that our model, in the limiting case of two
immiscible fluids, produces physically realistic interfacial tensions.
In terms of the basic two-species Rothman-Keller immiscible lattice-gas,
surface tension has been extensively investigated from both a
theoretical and a numerical viewpoint by Adler, d'Humi\`{e}res and
Rothman~\cite{bib:ahr}. Using a bubble experiment as described in their
paper, we can check the validity of our basic model by evaluating the
surface tension in the immiscible fluid case. We use Laplace's law,
which in two dimensions is
\bge
P_{\mbox{in}} - P_{\mbox{out}} = \frac{\sigma}{R},
\ee
where $R$ is the radius of the bubble, $P_{\mbox{in}}$ is the average
pressure within the bubble and $P_{\mbox{out}}$ the average pressure
outside. The results from our simulations are shown in
Fig.~\ref{fig:stb}. They give good agreement with Laplace's law, and a
best-fitting line through the origin results in an estimate of
$\sigma\approx 0.378$, close to the results reported by Adler and
co-workers~\cite{bib:ahr}.

We now turn to the analysis of the interfacial tension for varying
initial concentrations of surfactant in the system, where in each case
the amphiphile is added at the initial bulk oil-water interface. We make
use of a direct method of calculating the surface tension across this
interface, although, due to the complex dynamical nature of the
microemulsion behavior we are modelling, as more surfactant is added we
have to allow for progressively more extensive relaxation of the system
before collecting data on the equilibrated system.  For the case of a
flat interface perpendicular to the $z$-axis, the surface tension
$\sigma$ is given by the integral over $z$ of the difference between the
component $P_{N}$ of pressure normal to the interface and the component
$P_{T}$ transverse to the interface~\cite{bib:ahr}:
\bge
\sigma = \int_{-\infty}^{\infty} \left[ P_{N}(z) - P_{T}(z)
\right] dz .
\ee
This quantity can be calculated by empirically computing the integral
which, for a simulation cell of lattice size $L_{\perp}$ by
$L_{\parallel}$ and physical coordinates of lattice sites given by
$x_{\perp}$ and $x_{\parallel}$, takes the form
\bge
\sigma = \left< \frac{\sqrt{3}}{2L_{\parallel}} \sum_{i=0}^{6}
\sum_{\frac{L_{\perp}}{4} \leq x_{\perp} < \frac{3L_{\parallel}}{4}}
(c_{i\perp}^{2} - c_{i\parallel}^{2}) n_{i}(x_{\perp},x_{\parallel})
\right>
\label{eq:sdni}
\ee
where $c_{i\perp}$ and $c_{i\parallel}$ are the components of the
underlying lattice vectors $\bfc_{i}$ perpendicular and parallel to the
interface, $n_{i}(x_{\perp},x_{\parallel})$ is the total number of
particles present at a site and where the angular brackets denote an
average over time.
\begin{figure}
\begin{center}
\leavevmode
\hbox{%
\epsfxsize=3.5in
\epsffile{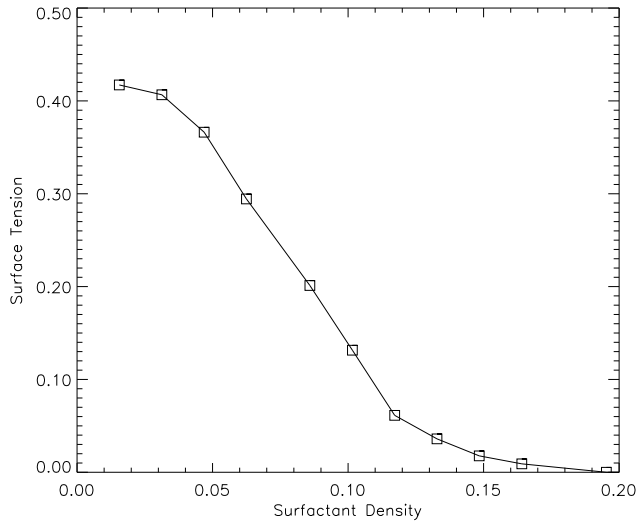}}
\end{center}
\caption{Surface tension as calculated for varying amounts of
amphiphile in the system.}
\label{fig:stva}
\end{figure}
There is an important caveat to be mentioned here: as the initial
concentration of surfactant in the system is increased one finds that
the initially stable oil-water interface begins to show signs of
distortion and break up. This effect arises as a result of the
energetics of the system; preferentially surfactant particles reside in
thin (mono)-layers at oil-water surfaces and consequently the system
acts to create as much oil-water interfacial length as possible in order
to accommodate the amphiphile. At some critical density of surfactant it
is clear that the initially flat interface will break up completely;
indeed, we then see the formation of a bicontinuous ``middle'' phase
corresponding to an effective oil-water surface tension of zero. At and
beyond this point the methods outlined above for measuring $\sigma$ are
ineffective. However, below this critical density we expect to be able
to use Eq.~(\ref{eq:sdni}) to evaluate the surface tension, bearing in
mind that, as we add more surfactant to the system, we have to wait
correspondingly longer for the interface to relax before relevant
measurements of $<n_{i}>$ can be made; we expect sections of negative
values for $\sigma$ prior to equilibration and denote these as being
part of a ``transient region''. In the asymptotic region the interface
stabilizes and only positive average values for the surface tension are
found; this is denoted as the ``smooth region.''

The results shown in Fig.~\ref{fig:stva} are the values of interfacial
tension obtained, {\it once the smooth region has been reached}, for
varying initial concentrations of surfactant in the system; the error
bars from the subsequent time average are smaller than the size of the
symbols and so are not included. The plotted values of surfactant
density are given as a proportion of the total reduced density of oil,
water and surfactant in the system~\cite{bib:bce}. The figure clearly
shows that our model is behaving as one would expect; the interfacial
tension is reduced dramatically by the presence of amphiphile in accord
with experiment~\cite{bib:agk}. It is worth mentioning that our results
also closely mimic the interfacial behavior in a mixture of two
immiscible polymers to which is added a linear diblock copolymer
comprised of units of both of the immiscible polymers: a sharp decrease
in interfacial tension is observed with the addition of a small amount
of copolymer, which compares well with the linear section of the graph,
followed by a levelling off as the copolymer concentration is increased.
The levelling off at higher concentrations is indicative of interfacial
saturation by the copolymer and subsequent formation of copolymer
micelles dispersed in the homopolymer phases. The relatively flat region
of the curve at very low surfactant densities is due to the tendency of
a certain number of amphiphiles to exist as highly dynamic monomers
within the bulk oil and water regions. As more surfactant is
subsequently added to the system, these molecules preferentially align
at the energetically favourable oil-water interfaces and so begin to
strongly influence the interfacial tension.

We find that, as the surfactant density is increased, the transient
regime persists for a longer time: $1000$ timesteps for a surfactant
density of $0.0156$, $3000$ timesteps for $0.0469$, and $13000$
timesteps for $0.1172$.  This implies that we are moving towards a {\em
  critical} value of the surfactant density at which the flat interface
will break up altogether and a smooth regime will never arise. This is
first seen in our simulations with a surfactant density of $\approx
0.195$; beyond this value, the computed average surface tension
remains negative over the entire simulation and permanent break-up of
the initially flat interface is observed. In Fig.~\ref{fig:stva} we
have designated this point as corresponding to an effective surface
tension of zero.

\section{Phase separation in binary immiscible fluids}
\label{sec:bif}

In the $2D$ binary oil-water limit of our model we expect the domain
growth exponent to be $n = \frac{2}{3}$, in line with the results of
previous lattice-gas and related models. This is consistent with being
in the inertial hydrodynamic regime, where the hydrodynamic length
$R_{h}$ is less than the domain size $R$, a condition forced on prior
lattice-gas models~\cite{bib:bal} as a result of their inability to vary
viscosity or surface tension independently of density.  A benefit of our
model is that we can access the other scaling regime, where $R < R_{h}$,
in a consistent manner. This is possible because of the presence of the
inverse {\it temperature-like} parameter $\beta$ that we have introduced
into our lattice-gas model (see Eq.~(\ref{eq:bd})). This gives us
exactly the desired form of control, since we are able to alter the
surface tension (and related viscosity) without having to change the
density and consequently we can reach the $R < R_{h}$ regime. In this
case we expect to find a growth exponent of $n = \frac{1}{2}$, resulting
from a droplet-coalescence mechanism~\cite{bib:smgg,bib:bal}, in
accordance with the predictions of Furakawa~\cite{bib:f} and the
molecular dynamics simulation results of Velasco and
Toxvaerd~\cite{bib:vat}. We note that this high-viscosity regime can
also be accessed by lattice-Boltzmann models and that one such model was
used recently for an investigation into binary fluid spinoidal
decomposition~\cite{bib:oosyb}. However, it should be noted that these
lattice-Boltzmann simulations do not include fluctuations; if such
features are regarded as desirable, the noise has to be inserted by hand.

\begin{figure}
\begin{center}
\leavevmode
\hbox{%
\epsfxsize=3.5in
\epsffile{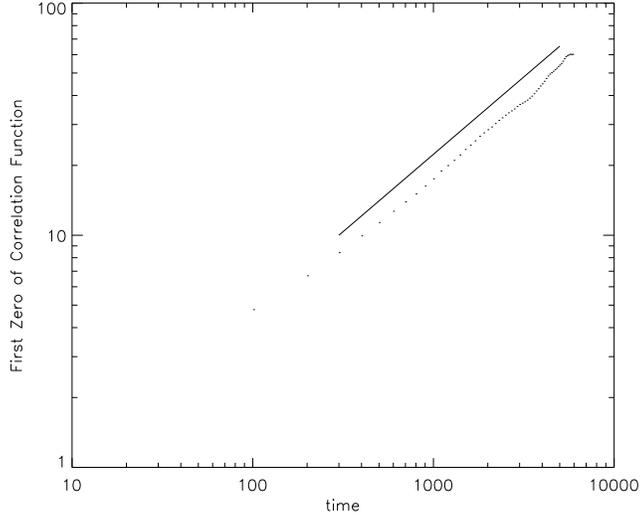}}
\end{center}
\caption{Temporal growth of domain size, $R(t)$, for binary fluid and
$\beta = 0.5$, shown in a logarithmic-scale plot. The straight line
has gradient $2/3$ and is included as a guide only.}
\label{fig:scahamow17}
\end{figure}

To analyze the domain growth quantitatively we obtain the first zero
crossing of the coordinate-space pair-correlation function, which is
equal to the characteristic domain size, $R(t)$. At time $t$ following
the quench, this correlation function is given by
\bge
C({\bf r}, t) = \frac{1}{V} < \sum_{\bf x} q({\bf x},t)
q({\bf x} + {\bf r}, t) >
\ee
where $q({\bf x},t)$ is the two-fluid (oil and water) density difference
(total color charge) at each site, $V$ is the volume, and the average is
taken over an ensemble of initial conditions. Taking the angular average
of $C({\bf r}, t)$ gives $C(r,t)$, the first zero crossing of which
gives a measure of the characteristic domain size. Typically, at least
five independent runs were averaged to determine the growth law for each
system studied.

\begin{figure}
\begin{center}
\leavevmode
\hbox{%
\epsfxsize=3.5in
\epsffile{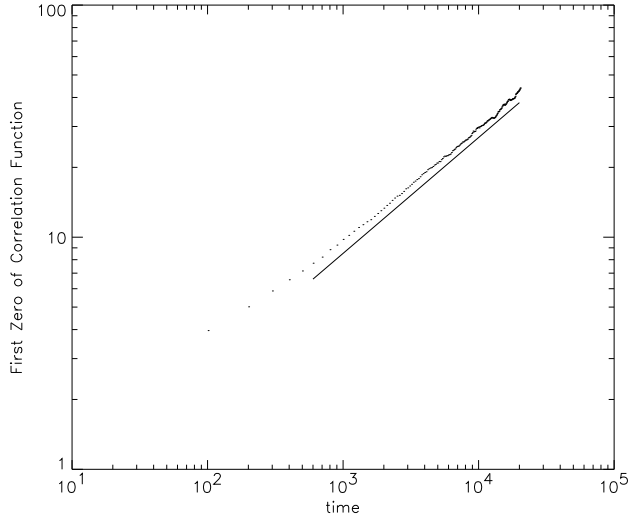}}
\end{center}
\caption{Temporal growth of domain size, $R(t)$, for binary fluid and
$\beta = 0.137$, shown in a logarithmic-scale plot. The straight line
has gradient $1/2$ and is included as a guide only.}
\label{fig:scahamow27}
\end{figure}

Setting $\alpha = 1.0$ in Eq.~(\ref{eq:tiw}) and the inverse
temperature-like parameter $\beta = 0.5$, we perform a critical quench
(that is, with equal amounts of oil and water in the system) on a
simulation cell of size $256 \times 256$. The initial condition is
random placement of the oil and water particles on the underlying
lattice. The result is shown in Fig.~\ref{fig:scahamow17}, where we have
used logarithmic scales so as to be able to observe any exponent in the
algebraic power law growth for the system. The domain growth exponent is
clearly $n = \frac{2}{3}$, consistent with previous results obtained for
the Rothman-Keller model~\cite{bib:bal} and characteristic of the regime
$R > R_{h}$. It is worth mentioning that, as expected, we obtain this
behavior for a wide range of values of $\beta$, from $0.3$ upwards.

Further lowering of $\beta$ and hence $\sigma$, the interfacial tension,
allows us to access the $R < R_{h}$ regime, where typically the domain
size is expected to be less than the hydrodynamic length. We do indeed
observe a different scaling exponent.  The result for $\beta = 0.137$ is
contained in Fig.~\ref{fig:scahamow27}, where the exponent is clearly $n
= \frac{1}{2}$. Although not included in this figure, at later times
than those shown or, alternatively, with larger values of $\beta$, we
have observed the beginnings of crossover to $t^{2/3}$ behavior,
consistent with expectations. The crossover from $t^{1/2}$ to
$t^{2/3}$ growth behaviour occurs at a progressively earlier time as
$\beta$ is systematically increased from $0.137$. Once $\beta$ gets
close to $0.3$ then the $n = \frac{1}{2}$ behaviour is no longer seen,
the crossover effect disappears and we get $n = \frac{2}{3}$ growth
right from the start of the simulations.

\section{Self-Assembly Dynamics in Microemulsions}
\label{sec:sadm}

\begin{figure}
\begin{center}
\leavevmode
\hbox{%
\epsfxsize=6.5in
\epsffile{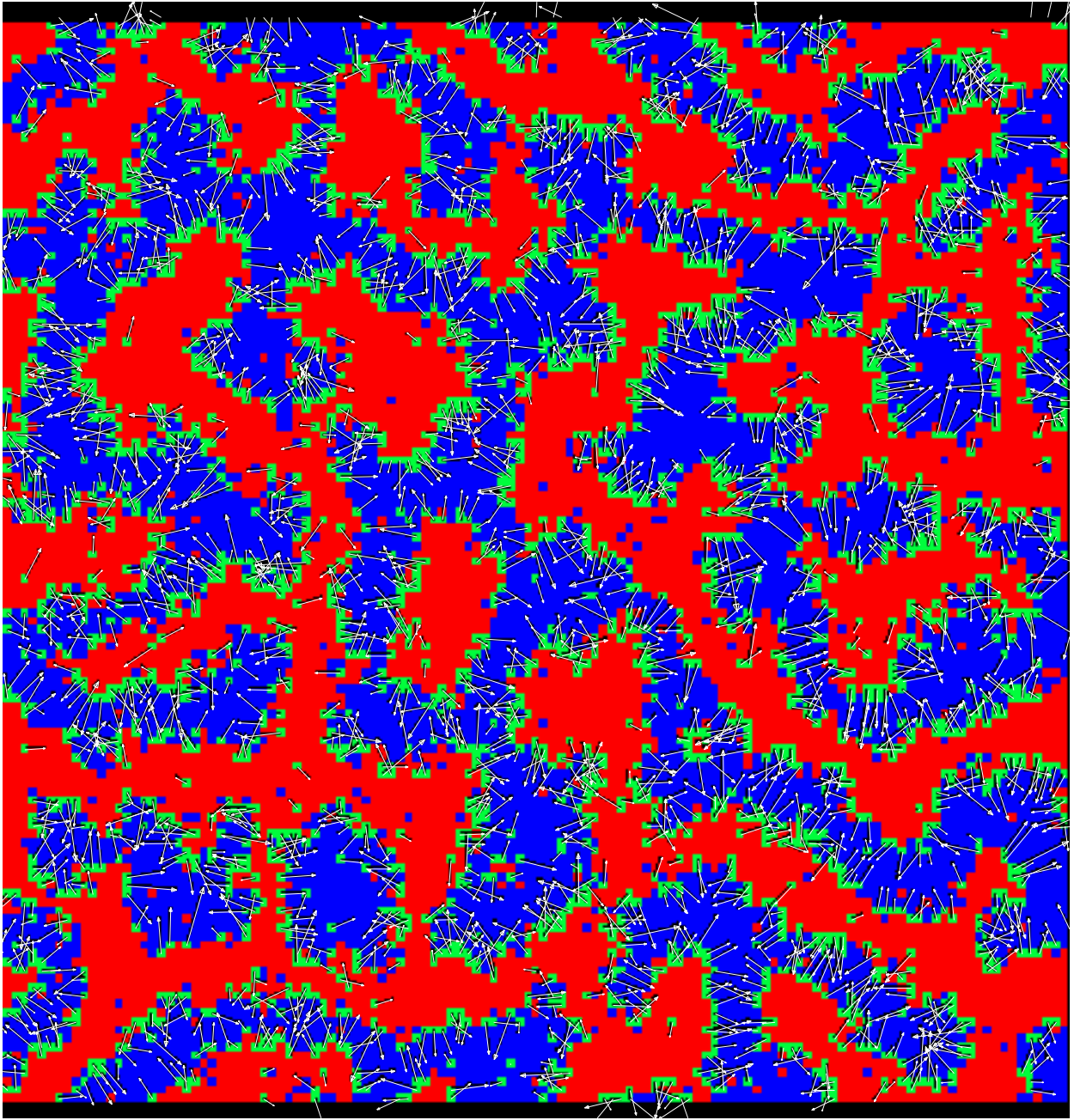}}
\end{center}
\caption{Bicontinuous microemulsion structure shown at timestep $200$
of a simulation with equal amounts of oil and water in the system. The
arrows depict the direction of the amphiphile vectors: note that they
always point from the oil to water domains as we expect.}
\label{fig:colps}
\end{figure}

We now turn to the analysis of the ternary system.  It is clear that the
presence of surfactant in an oil-water mixture dramatically alters the
interfacial energetics (in particular it lowers the interfacial tension)
and so it will affect the growth of domains and consequently alter the
usual binary-fluid scaling phenomena.  When there is sufficient
amphiphile present we expect to see some final characteristic domain
size $R_{c}$ imposed on the system as it reaches an equilibrium
state. The effect that the amphiphile molecules have on the usual
oil-water immiscible behaviour is clearly shown in
Fig.~\ref{fig:colps}, which depicts timestep $200$ of a simulation of
a bicontinuous microemulsion
phase; the arrows show the direction and size of the colour
dipole vectors which represent the surfactant particles. We note that
as expected the surfactant particles migrate to the oil-water
interfaces and always tend to point from one color to the other,
sugesting that they are exhibiting a hydrophilic-hydrophobic like
nature. Due to the immiscibility of the oil
and water particles, all oil-water
interfaces seek to contract in length as much as possible; the very
strong requirement that the surfactant particles sit at such interfaces,
however, means that at some point the shrinking must cease so the system
establishes its saturated domain size. The underlying lattice-gas
dynamics will of course still be present in such a system, but, by
averaging over an ensemble of simulations and over time we expect to be
able to determine $R_{c}$.

We begin with equal amounts of oil and water in our system while the
amount of surfactant is varied for each simulation. We note that this
leads to growth of bicontinuous as opposed to droplet phases and that
these are effectively equivalent to the critical quenches investigated
in the binary-fluid case. Again we work with $\beta = 1.0$ and use
\bgeas
  \alpha &=& 1.0  \\
     \mu &=& 0.001\\
\epsilon &=& 8.0  \\
   \zeta &=& 0.005
\eeas
as the values of the coupling coefficients in Eq.~(\ref{eq:tiw}). In
essence this choice requires the surfactant particles to sit at the bulk
oil-water interfaces and discourages the formation of micelles which
would hamper the consistent measure of the characteristic length scale
of the bicontinuous domain.  The results that follow for the ternary
system have been obtained on a $128 \times 128$ lattice with periodic
boundary conditions in both ($x$ and $y$) directions, and with the
particles initially placed on the lattice at random. The amount of
surfactant used in each simulation is given in terms of its reduced
density; the amount of oil and water in the system is kept constant,
being at a reduced density of $0.17$ for each simulation. The
measurement of the domain size $R(t)$ is calculated from the spatial
pair-correlation function, as described in Sec.~\ref{sec:bif}.

\begin{figure}
\begin{center}
\leavevmode
\hbox{%
\epsfxsize=3.5in
\epsffile{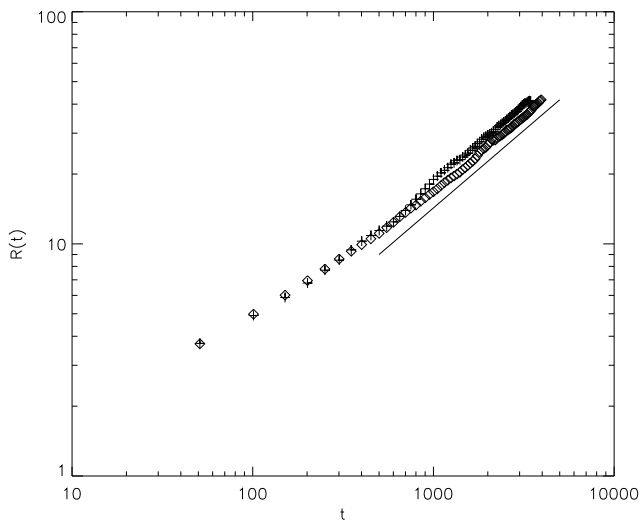}}
\end{center}
\caption{Temporal growth of domain size, shown in a logarithmic-scale
  plot. The straight line has gradient $2/3$ and is included as a guide
  to the eye. The upper symbols (crosses) are for $0.02$ surfactant and
  the lower diamonds are for $0.04$ surfactant.}
\label{fig:scaosw3839}
\end{figure}

The results for systems with reduced surfactant densities of $0.02$ and
$0.04$ are shown in Fig.~\ref{fig:scaosw3839}. For the former an average
over five simulations is shown, while the latter consists of an average
over ten. Over the late-time scaling regime, domain growth in both of
these systems clearly proceeds with an algebraic exponent of $n =
\frac{2}{3}$. There is insufficient amphiphile in the system to affect
the oil-water binary immiscible fluid behavior. As described in
Sec.~\ref{sec:sta}, this is consistent with the expected presence of a
certain number of background amphiphilic monomers within the bulk oil
and water regions, an effect which in reality is dependent on the
strength and type of the amphiphile employed. If there is any change in
domain growth due to the tiny amount of surfactant present in these two
simulations, it would only be observed at very late times on
significantly larger lattices than those we have used here.  We can
investigate such effects, however, by simply starting with more
surfactant in the system.  From our analysis in Sec.~\ref{sec:sta} we
expect a significant reduction in the surface tension to occur at the
equivalent of a reduced density of $\simeq 0.05$ surfactant and beyond.
At this point, large numbers of surfactant molecules have attached
themselves to the oil-water interfaces and so begin to affect the
dynamical growth of domains. Although not shown here, with a reduced
density of $0.05$ surfactant in the system we obtain a crossover from an
exponent $n = \frac{2}{3}$ to $n = \frac{1}{2}$ at late times as
surfactant molecules adsorb at the interfaces and, as expected, begin to
affect the domain growth.

\begin{figure}
\begin{center}
\leavevmode
\hbox{%
\epsfxsize=3.5in
\epsffile{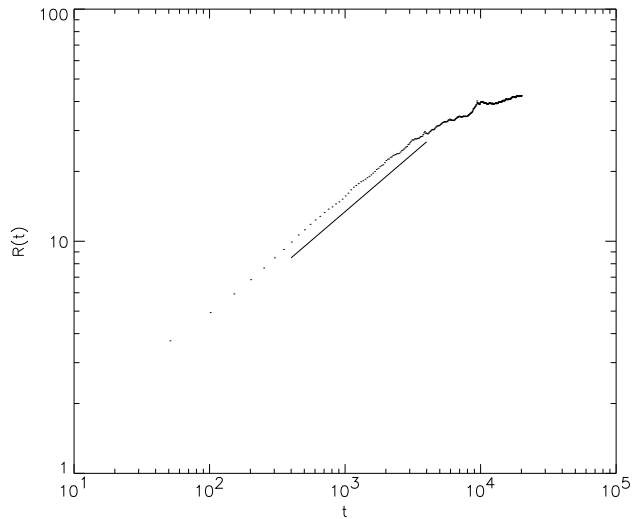}}
\end{center}
\caption{Temporal growth of domain size, shown in a logarithmic-scale
  plot. The straight line has gradient $1/2$ and is included as a guide
  to the eye. Here we have $0.06$ surfactant.}
\label{fig:scaosw37}
\end{figure}

With reduced surfactant densities of $0.06$ (and $0.07$), we observe a
growth exponent of $n = \frac{1}{2}$ for a majority of the time
evolution, but in addition there is now a clear crossover to
slower-than-algebraic growth at late times. This is depicted in
Fig.~\ref{fig:scaosw37}, which contains the result for domain growth in
a system with $0.06$ surfactant and is obtained from an average over
nine simulations, each having different initial random number seeds. The
behavior described is not due to finite-size effects in the system, as
we have stopped the simulations well before this becomes a problem.  The
observed ``jump'' of the growth exponent from $n = \frac{2}{3}$ to $n =
\frac{1}{2}$ and then to slower behavior as the surfactant density is
increased is consistent with the binary-fluid behavior results that we
outlined in Sec.~\ref{sec:bif}. The drop in surface tension takes us
into a regime that is equivalent to the slow binary one and beyond this
to slower-than-algebraic growth. These results also show clear evidence
of the crossover scaling transition, alluded to by Laradji {\it et
  al.}~\cite{bib:lmtz}, from algebraic binary growth ($n = \frac{1}{2}$)
to a slower domain growth when surfactants are present.  Our use of a
lattice-gas model, in contrast to the molecular dynamics technique
employed by these authors, has the advantage of easy access to a wide
range of different timescale regimes, as the results we obtain here make
evident.

\begin{figure}
\begin{center}
\leavevmode
\hbox{%
\epsfxsize=3.5in
\epsffile{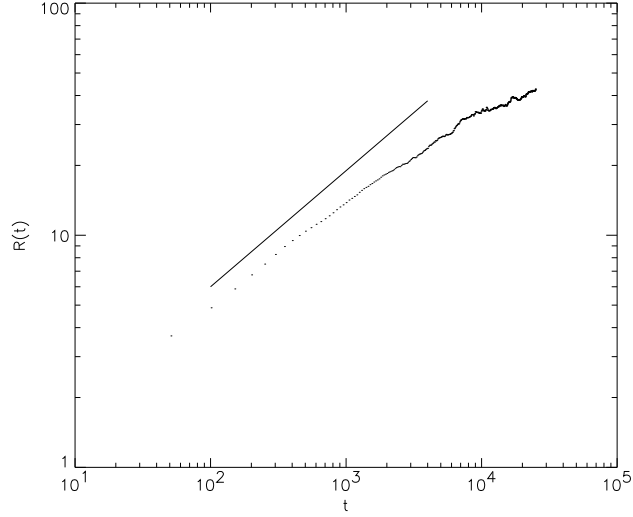}}
\end{center}
\caption{Temporal growth of domain size, shown in a logarithmic-scale
  plot. The straight line has gradient $1/2$ and is included as a guide
  to the eye. Here we have $0.08$ surfactant.}
\label{fig:scaosw36}
\end{figure}

\begin{figure}
\begin{center}
\leavevmode
\hbox{%
\epsfxsize=3.5in
\epsffile{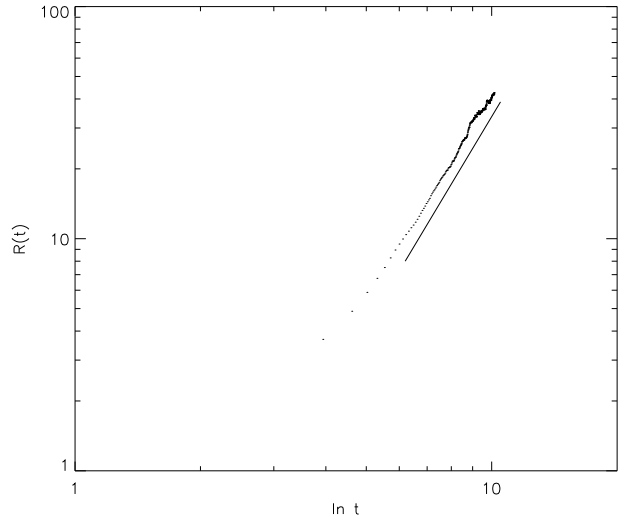}}
\end{center}
\caption{Plot of $\ln t$ against growth of domain size, shown with
  logarithmic scales and surfactant density of $0.08$. The straight line
  has gradient $3.0$ and is included as a guide only.}
\label{fig:scaosw36ln}
\end{figure}

Increasing the initial reduced density of amphiphile to $0.08$ we see a
clear departure from algebraic behavior over the timescale of the
simulations: After the first $400$ timesteps the slope of the curve is
consistently below the line $n = \frac{1}{2}$, as shown in
Fig.~\ref{fig:scaosw36}. Consequently we look at a plot of $\ln t$
against domain size in order to investigate whether we now have
logarithmically slow, or just slow, growth in this region. As before
this is shown plotted on logarithmic scales (see
Fig.~\ref{fig:scaosw36ln}) so that we are able to observe any
algebraic exponent for the $\ln t$ growth.  If the 
slow growth in these systems can indeed be related in some way to that
in systems with quenched impurities~\cite{bib:ghss}, then we would
expect to find some power $\theta$ for the growth function $(\ln
t)^{\theta}$, which decreases as the amount of surfactant in the system
is further increased. In this initial case we find a value $\theta
\simeq 3.0$ for the timescale of the simulation beyond the very
early-time transient behavior.

\begin{figure}
\begin{center}
\leavevmode
\hbox{%
\epsfxsize=3.5in
\epsffile{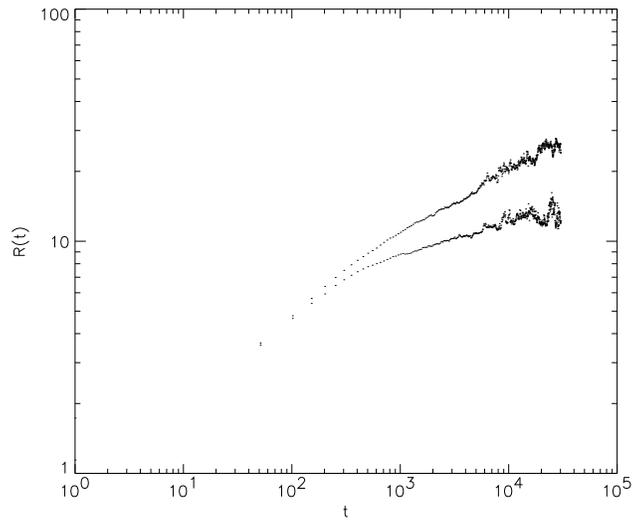}}
\end{center}
\caption{Temporal growth of domain size, shown in a logarithmic-scale
  plot. The upper points correspond to $0.10$ surfactant and the lower
  ones to $0.12$ surfactant.}
\label{fig:scaosw3235}
\end{figure}

\begin{figure}
\begin{center}
\leavevmode
\hbox{%
\epsfxsize=3.5in
\epsffile{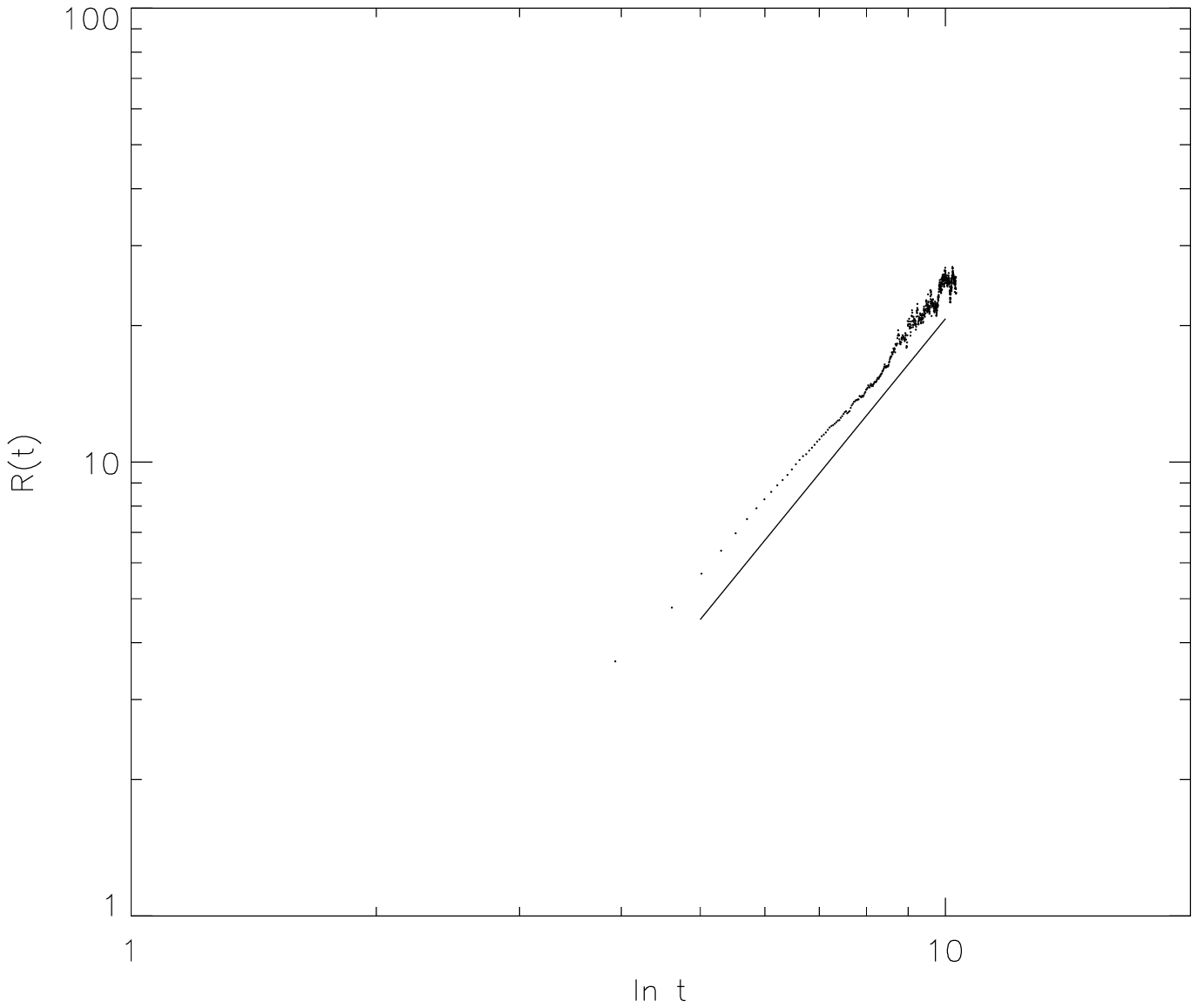}}
\end{center}
\caption{Plot of domain size against $\ln t$, shown with
  logarithmic scales and surfactant density of $0.10$. The straight line
  has gradient $2.2$ and is included as a guide only.}
\label{fig:scaosw32ln}
\end{figure}

\begin{figure}
\begin{center}
\leavevmode
\hbox{%
\epsfxsize=3.5in
\epsffile{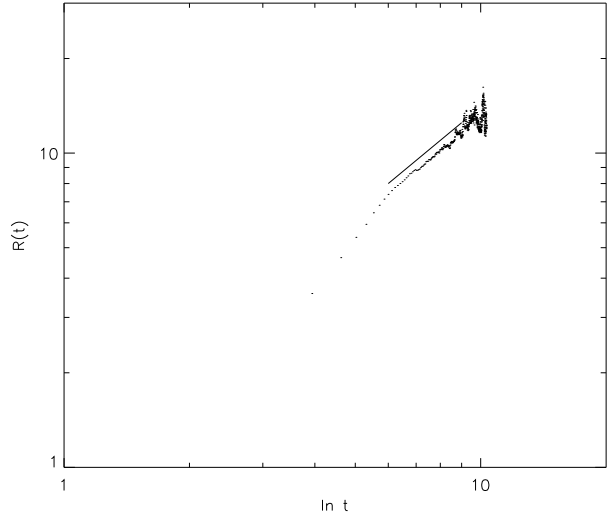}}
\end{center}
\caption{Plot of domain size against $\ln t$, shown with
  logarithmic scales and surfactant density of $0.12$. The straight line
  has gradient $1.1$, and is included as a guide only.}
\label{fig:scaosw35ln}
\end{figure}

Moving to simulations with higher quantities of amphiphile, it is clear
that there is enough surfactant present in the system for the domain
growth to be significantly retarded. Fig.~\ref{fig:scaosw3235} contains
logarithmic-scale plots of $R(t)$ versus $t$ for $0.10$ and $0.12$
reduced density of surfactant and shows that we are now in a regime
where we get complete cessation of domain growth well within the
finite-size limits of the system. The former of these is the result of
an average over fourteen runs, and the latter an average over ten. In a
similar fashion as above we re-analyze these two results, again using
logarithmic-scale plots of $R(t)$ versus $\ln t$ in order to establish
whether a value for the exponent $\theta$ can be extracted to help
clarify the nature of the ``slow'' growth observed.
Fig.~\ref{fig:scaosw32ln} contains a logarithmic-scale plot of $R(t)$
versus $\ln t$ for the first of these ($0.10$ surfactant) and shows slow
logarithmic growth with exponent $\theta \simeq 2.2$.  With surfactant
densities of $0.12$ and higher it is clear from the logarithmic-scale
plots of $R(t)$ versus $t$ that a significant slowing-down occurs after
approximately the first $400$ timesteps of the simulations (designated
as the transient region).  This is obviously related to the time
required for a significant proportion of the surfactant molecules
present to migrate to the oil-water interfaces that form rapidly at very
early times.  Consequently we look at later times to establish a value
for the exponent $\theta$.  Fig.~\ref{fig:scaosw35ln}, again a
logarithmic-scale plot of $R(t)$ versus $\ln t$, but in this case for a
surfactant density of $0.12$, gives an approximate exponent of $\theta
\simeq 1.1$ over the majority of the simulation running time, followed
by saturation of the domain size at late times.  With these intermediate
surfactant densities, as clearly shown in Fig.~\ref{fig:scaosw3235}, we
observe large fluctuations in the measured domain size at late times in
the simulations which cannot be eliminated by ensemble averaging.
Indeed, these fluctuations have an important physical basis in that they
correspond to the continual break up and reformation of the
bicontinuous-like structures under investigation, resulting from the
finely balanced competition between the immiscible binary-fluid behavior
of oil and water and the action of surfactant molecules at oil-water
interfaces. As we increase the density of surfactant beyond this level,
we find that the fluctuations become less severe and actually die out
because sufficient surfactant molecules reside at the interfaces to
effectively outweigh the oil-water interfacial tension completely.  The
domain structures then become strongly pinned and consequently less
fluctuation is allowed by the system.

\begin{figure}
\begin{center}
  \leavevmode
\hbox{%
  \epsfxsize=3.5in \epsffile{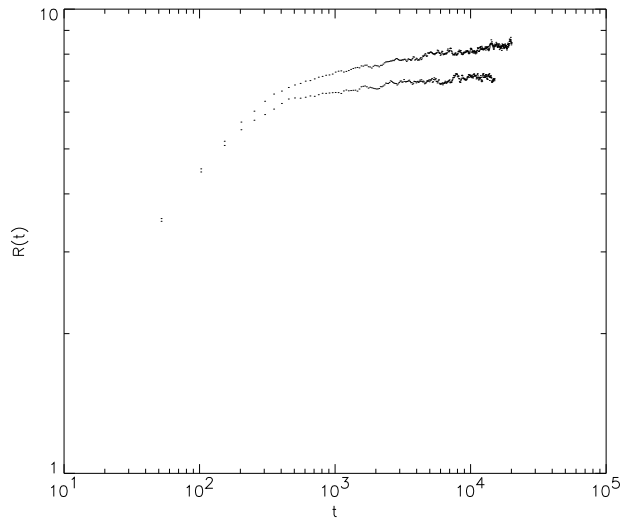}}
\end{center}
\caption{Temporal growth of domain size, shown in a logarithmic-scale
  plot. Moving from top to bottom the points correspond to $0.14$, and
  $0.15$ surfactant respectively. The upper curve is an average over ten
  simulations, the lower five.}
\label{fig:scaosw3441}
\end{figure}

\begin{figure}
\begin{center}
\leavevmode
\hbox{%
\epsfxsize=3.5in
\epsffile{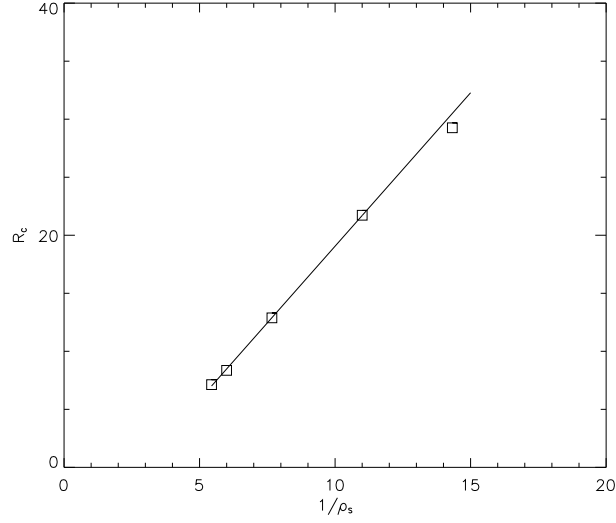}}
\end{center}
\caption{Plot of the average final characteristic domain size $R_{c}$
  against the inverse of the density of surfactant $1/\rho_{s}$ at the
  interfaces in the system.}
\label{fig:fcsod}
\end{figure}

Fig.~\ref{fig:scaosw3441}, which contains logarithmic-scale plots of
$R(t)$ versus $t$ for simulations with relatively high amounts of
amphiphile, clearly shows that the domain growth is finally halted by
the presence of sufficient surfactant: In essence we obtain a final
characteristic saturated domain size for the equilibrium structures
formed by the system. We expect that the average domain size will stop
growing when all of the oil-water interface is covered by a surfactant
``monolayer'' \cite{bib:lmtz}. Noting that the average domain size,
$R(t)$, is inversely proportional to the total length of such oil-water
interfaces, we then expect the final domain size to be inversely
proportional to the average density of surfactants at these interfaces.
However, in contrast to the deep quenches with no system fluctuations
performed by Laradji {\it et al.}~\cite{bib:lmtz}, where all the
surfactant molecules are found at oil-water interfaces, we have a
situation wherein a certain amount of the surfactant is likely to exist
as monomer in bulk oil and water regions, this being confirmed by our
surface tension analysis (see Sec.~\ref{sec:sta}). Consequently, in
plotting the final domain size $R_{c}$ as a function of $1/\rho_{s}$,
where $\rho_{s}$ is the average density of surfactant at the oil-water
interfaces, we need to evaluate $\rho_{s}$ from the total amount of
surfactant in a particular system by subtracting away the ``background
monomer density.''  The result is plotted in Fig.~\ref{fig:fcsod}: We
find the expected linear relationship between the final saturated domain
size and the amount of interfacial surfactant in the system; that is,
the final characteristic domain size is inversely proportional to the
{\it interfacial} surfactant density in the system. The straight line on
the plot is a linear fit to the first four points. (The final point,
corresponding to a total reduced surfactant density of $0.09$, lies
below this line probably because the simulation had not fully
equilibrated.) It is worth noting that the result shown in
Fig.~\ref{fig:fcsod} is also consistent with the relationship found
between the final domain size and the amplitude of disorder in systems
with quenched impurities, as determined by Gyure {\it et
al.}~\cite{bib:ghss}.

\begin{figure}
\begin{center}
\leavevmode
\hbox{%
\epsfxsize=3.5in
\epsffile{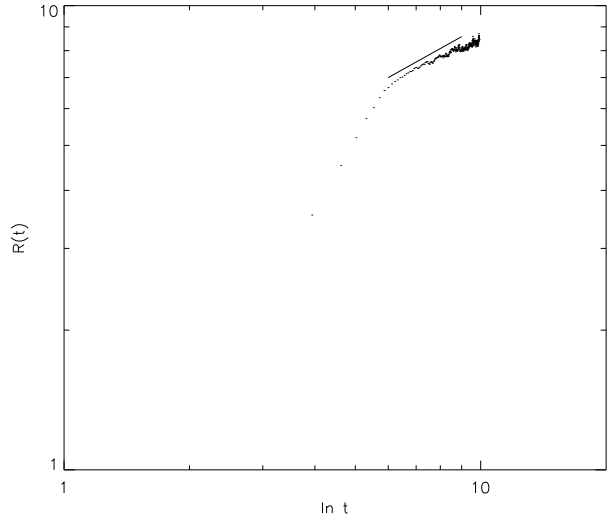}}
\end{center}
\caption{Plot of domain size against $\ln t$, shown with
  logarithmic scales and surfactant density of $0.14$. The straight line
  has gradient $0.5$ and is included as a guide only.}
\label{fig:scaosw34ln}
\end{figure}

\begin{figure}
\begin{center}
\leavevmode
\hbox{%
\epsfxsize=3.5in
\epsffile{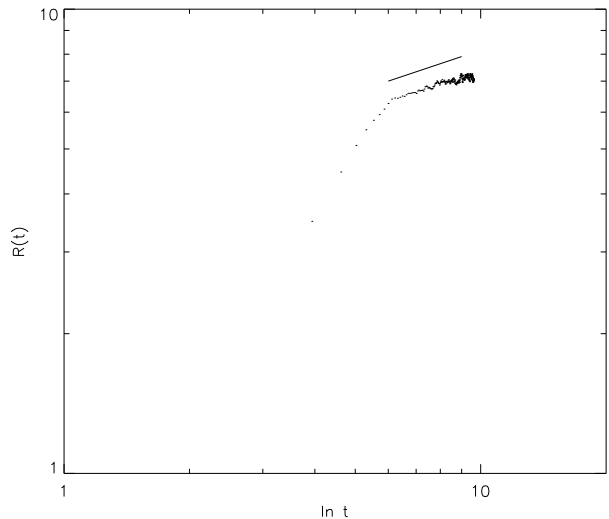}}
\end{center}
\caption{Plot of domain size against $\ln t$, shown with
  logarithmic scales and surfactant density of $0.15$. The straight line
  has gradient $0.3$ and is included as a guide only.}
\label{fig:scaosw41ln}
\end{figure}

Fig.~\ref{fig:scaosw34ln}, which is a plot of domain size versus $\ln t$
for the case of surfactant density $0.14$, indicates that in this case
the slow domain growth may go as $(\ln t)^{\theta}$ with $\theta \simeq
0.5$ over the dominant timescale of the simulations (beyond the initial
transient region) and before the domain size saturates completely. The
same is true for $0.15$ surfactant but in this case $\theta \simeq 0.3$
before saturation occurs (see Fig.~\ref{fig:scaosw41ln}).
Table~\ref{tab:theta} contains a summary of how the exponent $\theta$
varies with surfactant concentration in the region of logarithmically
slow growth studied in these and the previous simulations.

\begin{figure} 
\begin{center}
\begin{tabular}{|c|c|} \hline
\em{Surfactant Concentration}    &  \em{$\theta$} \\ \hline
$0.08$  &  $3.0$  \\
$0.10$  &  $2.2$  \\
$0.12$  &  $1.1$  \\
$0.14$  &  $0.5$  \\
$0.15$  &  $0.3$  \\ \hline
\end{tabular}
\end{center}
\caption{Logarithmic exponent $\theta$ as it changes with
surfactant concentration}
\label{tab:theta}
\end{figure}

These results are consistent with the picture obtained from analysis of
domain growth with quenched impurities, where the slow growth goes as
$(\ln t)^{\theta}$, and where $\theta$ changing as the number of
impurities is increased~\cite{bib:ghss}.

\section{Discussion and Conclusions}
\label{sec:dac}

We have studied both binary immiscible and ternary microemulsion
dynamical behavior using our hydrodynamic lattice-gas model of
self-assembling amphiphilic systems. In the binary case we have found
algebraic scaling laws in agreement with expectations~\cite{bib:bray}, 
the $2D$ growth exponents being
$\frac{1}{2}$ and $\frac{2}{3}$ at early and late times respectively.
The former is
new to lattice-gas models, although it has also been observed in
molecular~\cite{bib:vat}, Langevin~\cite{bib:lwac} and dissipative
particle dynamics~\cite{bib:cn}
simulations, and is also in accord with the results of a
renormalization-group approach~\cite{bib:bray}. 
In the ternary system we have confirmed, in
accord with experiment, that the presence of surfactant results in a
reduction of the oil-water interfacial tension and consequently that the
growth of domains in such systems is radically different from growth in
the binary case. We find a crossover from the fast $n = \frac{2}{3}$
binary regime in which we begin, first to $n = \frac{1}{2}$ algebraic
growth and then to ``slow'' behavior as surfactant is added to the
system. This behavior mimics exactly the crossover scaling function
predicted by Laradji {\it et al.}~\cite{bib:lmtz} from molecular dynamic
simulations of similar systems. The greater the concentration of
surfactant the slower the growth becomes; in fact, it appears to be
logarithmically slow, the domain size going as $(\ln t)^{\theta}$ with
$\theta$ changing from $3.0$ through to $0.3$ as the amphiphile
concentration increases through the range considered in this study. This
behavior can be related to that of systems with quenched impurities in
which the domains are pinned at late times, although it is not presently
clear whether the logarithmic growth behavior observed is understandable
on this basis alone. Our lattice-gas model also enables us to access the
asymptotic, late-time regime in which the average domain size becomes
saturated. This occurs for intermediate to high surfactant
concentrations; we find the expected physical relationship between the
final characteristic domain size and the inverse of the interfacial
surfactant density.

In conclusion, we have completed an investigation into the complex
dynamical behavior of the two-dimensional bicontinuous microemulsion
phase, which corresponds to a critical quench in a binary oil and water
system. However, our model is also able to accurately simulate
off-critical droplet and micellar phases~\cite{bib:bce} and further work
is required to unravel the domain growth dynamics in such situations; we
expect the dynamical growth laws to be modified in some way since this
is also the case for the related binary-fluid off-critical quench.
Although the work reported in this paper has all been done in two
spatial dimensions, we are currently implementing a three-dimensional
version of our model~\cite{bib:toappear} where again, since binary
growth laws are different in three dimensions, we expect our present
results to be modified accordingly.

\section*{Ackowledgments}

ANE and PVC thank Mike Swift, Julia Yeomans, Enzo Orlandini and
Giuseppe Gonella for numerous stimulating discussions, and Ruta
Devalia for help with the color graphics.
BMB thanks Gene Stanley, Steve
Harrington, and Francis Starr for pointing out the existence and
relevance of one of the references~\cite{bib:ghss}, and for helpful
discussions. ANE is grateful to EPSRC and Schlumberger Cambridge
Research for funding his
research. PVC and BMB are indebted to NATO for partial support for this
project. BMB is supported in part by Phillips Laboratory and by the United
States Air Force Office of Scientific Research under grant number
F49620-95-1-0285.

\end{document}